\documentclass[amsmath,amssymb,aps,showkeys,showpacs,twocolumn,prl]{revtex4-1}

\usepackage[english]{babel}
\usepackage{graphicx}
\usepackage{graphics}
\usepackage{amsmath}
\usepackage{dcolumn}
\usepackage{amssymb}
\usepackage{bm}
\usepackage[latin1]{inputenc}
\usepackage{graphicx,color}
\usepackage[caption=false]{subfig} 
 
\begin{document}
\title{Magnetic oscillations for neutral atoms subject to an electromagnetic field }
\author{B. Farias}
\email{bruno.farias@ufcg.edu.br}
\affiliation{Centro de Ci\^encias e Tecnologia Agroalimentar, Universidade Federal de Campina Grande, 58840-000, Pombal, PB, Brazil.}
\author{C. Furtado}
\email{furtado@fisica.ufpb.br} 
\affiliation{Departamento de F\'isica, Universidade Federal da Para\'iba, Caixa Postal 5008, 58051-970, Jo\~ao Pessoa, PB, Brazil.} 

%
\begin{abstract}

We show that the de Haas van Alphen effect can be induced in a two dimensional atomic gas by the He-McKellar-Wilkens interaction mediated via an electric dipole moment. Under an appropriate field-dipole configuration, we show that the neutral atoms subject to a synthetic magnetic field arrange themselves in Landau levels. An experimental arrangement for observation of the atomic analog of dHvA oscillations is proposed. In a strong effective magnetic field regime we present the quantum oscillations in the energy and effective magnetization of the two dimensional atomic gas. From the dHvA period we determine the area of the Fermi circle of the atomic cloud.

\end{abstract}
\maketitle
\section{INTRODUCTION}
\label{intro}

It has been widely acknowledged that neutral atomic gases provide an ideal platform to experimentally investigate some fundamental phenomena originally connected with condensed matter systems \cite{1,2,3,4}. In particular, artificial magnetism for neutral atoms is attracting growing attention \cite{5,6,7,8,9}. In these systems, atoms interacting with a suitable configuration either of electric \cite{10}, magnetic \cite{11} or laser \cite{12} fields behave themselves as charged particles in presence of a magnetic field. The quest for artificial magnetism is to realize situations where a neutral particle acquires a geometrical phase when it follows a closed contour. 

In this context, in 1993, X.G. He and B.H.J. McKellar \cite{13} predicted the existence of a topological phase when an electric dipole encircles a line of magnetic monopoles. Due to the inherent difficulty in arranging a line of magnetic monopoles the idea of the original work of He and McKellar seemed purely speculative. However, in 1994, M. Wilkens \cite{14} proposed an experimental test, which considered the case of an electric dipole interacting with a magnetic field generated with ferromagnetic materials. Subsequently, Wei, Han and Wei \cite{15} introduced a more realistic test of the He-McKellar-Wilkens (HMW) phase for a neutral particle (with no permanent electric
dipole moment) that moves in an area where both a radial electric field and a uniform magnetic field are applied. The HMW phase has recently been experimentally measured using an atom interferometer, using a field geometry equivalent to that proposed by Wei, Han and Wei \cite{16}.

The motion of neutral particles in a plane and in the presence of a perpendicular synthetic magnetic field is highly special. This is because the cyclotron motion of the particles leads to the Landau level band structure \cite{17,18,18b,19}, with each level providing a macroscopic number of one-particle states that are strictly degenerate in energy. In this setting, the atomic gas can display an oscillatory dependence on physical observables as a function of field strength. Generally known as "quantum magnetic oscillations", these effects could be observed for example for the atomic analog of the magnetization (de Haas-van Alphen oscillations) \cite{20}, the resistivity (Shubnikov-de Haas oscillations) \cite{21}, the Hall resistance \cite{22}, or the specific heat \cite{23}.

In this work, we show that de Haas van Alphen (dHvA) oscillations can be induced in a two dimensional (2D) atomic system composed by neutral particles (with no permanent electric dipole moment) using a field-dipole configuration similar to the ones proposed by Wei, Han and Wei. Firstly, we demonstrate that the field-dipole configuration presented here makes the single-particle energy states of the atoms organize into Landau levels. In addition, we show that the confinement of the neutral particles in an atomic cloud restricts the magnetic field strength. Then, we propose an experimental configuration to realize the dHvA effect and we provide a simple qualitative picture of these oscillations. Finally, we estimate the area of the Fermi circle of the atomic cloud.

\section{LANDAU QUANTIZATION FOR NEUTRAL ATOMS IN PRESENCE OF AN ELECTROMAGNETIC FIELD}

We consider a 2D atomic cloud with each neutral particle of mass $M$ moving with a velocity ${\bf v}$ in an area where both a radial electric field and a uniform magnetic field are applied. 

For our purposes, the electric field (in cylindrical coordinates) has the following form
\begin{equation}\label{Eq1}
{\bf E} = \frac{\rho_{0}}{2 \epsilon_{0}} r \hat{r},
\end{equation} 
(where $\epsilon_{0}$ is the electric vacuum permittivity and $\rho_{0}$ is an uniform volume charge density). 

The magnetic field is applied along of $z$ and is given by
\begin{equation}\label{Eq2}
{\bf B} = B_{0} \hat{z}.
\end{equation}

These crossed fields polarizes the particle and consequently induces in it an electrical dipole given by
\begin{equation}\label{}
{\bf d} = \alpha \left( {\bf E} + {\bf v} \times {\bf B}\right),
\end{equation}
where $\alpha$ is the polarizability.

In this way the Lagrangian of the atom may be expressed as
\begin{equation}\label{}
\mathcal{L} = \frac{1}{2} M {\bf v}^2 + \frac{1}{2} \alpha \left( {\bf E} + {\bf v} \times {\bf B}\right)^2,
\end{equation}

Since the motion of the particle is constrained in the plane, i.e., ${\bf v} \bot {\bf B}$ the equation above can be rewritten as
\begin{equation}\label{}
\mathcal{L} =  \frac{1}{2} \left( M + \alpha {\bf B}^2 \right) {\bf v}^2 + \frac{1}{2} \alpha {\bf E}^2 + {\bf v} \cdot \left({\bf B} \times \alpha {\bf E} \right),
\end{equation}

The term ${\bf v} \cdot \left({\bf B} \times \alpha {\bf E} \right)$ is known as the energy of R\"otgen and is responsible for the modification in the canonical momentum of the system that differs from the mechanical component $M{\bf v}$.  In this way, we can write the Hamiltonian associated with this system as
\begin{equation}\label{Hamiltonian}
\mathcal{H} =  \frac{1}{2 m} \left[{\bf P} -  \left({\bf B} \times {\alpha\bf E}\right) \right]^2 - \frac{1}{2} \alpha {\bf E}^2,
\end{equation}
where $m = M + \alpha B_{0}^{2} $. Note that the interaction between the electromagnetic field and the electric dipole of the atom in the nonrelativistic limit coincides formally with the minimal coupling of a charged particle with an external magnetic field. As a consequence, we can define an effective vector potential for this system in the following way
\begin{equation}\label{}
{\bf A}_{\mathrm{eff}} =  \left(\alpha{\bf E} \times {\bf B}\right).
\end{equation}

According to the field configuration of Eqs. (\ref{Eq1}) and (\ref{Eq2}) we can rewrite
\begin{equation}\label{}
{\bf A}_{\mathrm{eff}} =  \frac{\alpha B_{0} \rho_{0}}{2 \epsilon_{0}} r \hat{\phi}.
\end{equation}

This effective vector potential corresponds to a constant magnetic field in the direction of z axis, since we have the relation
\begin{equation}\label{magnetic_field}
{\bf B}_\mathrm{eff} =  \nabla \times {\bf A}_{\mathrm{eff}} = \frac{\alpha B_{0} \rho_{0}}{\epsilon_{0}} \hat{z}.
\end{equation}

In order to obtain the atomic analog of the Landau levels, we will solve the Schrödinger equation associated to Eq. (\ref{Hamiltonian}). For this, we write the Schrödinger equation, in cylindrical coordinates, in the following form
\begin{eqnarray}\label{1}
-\frac{\hbar^{2}}{2 m} \bigg[\frac{1}{r} \frac{\partial}{\partial r} \left(r \frac{\partial}{\partial r} \right) &&+ \frac{1}{r^2} \frac{\partial^2}{\partial \phi^2} \bigg] \Psi + \bigg[- \frac{i\hbar \omega}{2}\frac{\partial } {\partial \phi} 
\nonumber\\&& + \frac{m \omega^{2}}{8} r^{2} 
+ \frac{m^{2} \omega^{2}}{8 \alpha B_{0}^2} r^{2} \bigg] \Psi = E \Psi,
\end{eqnarray}
with the cyclotron frequency
\begin{equation}
\omega = \frac{\alpha \rho_{0} B_{0}}{m \epsilon_{0}},
\end{equation} 

Once the coefficients in the differential Eq. (\ref{1}) are independent of the azimuth coordinate $\phi$, the angular momentum $ \hat{L}_{z} = i\hbar \frac{\partial}{\partial \phi}$ is a quantum integral of motion and so the wave function $\Psi$ can be factorized to separate the variables
\begin{equation}\label{2}
\Psi =  e^{i \ell \phi} R(r). 
\end{equation} 

Here $e^{i \ell \phi}$ is the eigenfunction of the operator $ \hat{L}_{z}$, with the eigenvalue $\hbar \ell$ where $\ell$ is an integer.

Substituting the solution (\ref{2}) into Eq.(\ref{1}) the Schr\"odinger assumes the form 
\begin{eqnarray}\label{3}
\frac{\hbar^{2}}{2 m} \left[\frac{d^2}{d r^2} + \frac{1}{r} \frac{d}{dr} - \frac{\ell^2}{r^{2}} \right] R 
+ \bigg[&& E - \frac{\hbar \omega \ell}{2}  - \frac{m \omega^{2}}{8} r^{2} \nonumber\\&&
- \frac{m^{2} \omega^{2}}{8 \alpha B_{0}^2} r^{2}  \bigg] R = 0.
\end{eqnarray}

Introducing the dimensionless variable $\xi = \frac{m \omega}{2 \hbar} r^2$ we rewrite Eq. (\ref{3}) in a dimensionless form
\begin{eqnarray}\label{4}
\left[\xi \frac{d^2}{d\xi^2} + \frac{d}{d\xi}\right] R + \left[- \frac{\ell^2}{4\xi} + \beta - \lambda \frac{\xi}{4} \right] R = 0,
\end{eqnarray} 
where $\beta = \frac{E}{\hbar \omega} - \frac{\ell}{2} $ and $ \lambda = 1 + \frac{m}{\alpha B_{0}^2}$.

The asymptotic analysis of Eq. (\ref{4}) prompts us to write a solution for $R(\xi)$ in the form
\begin{equation}\label{5}
R(\xi) = e^{-\lambda\xi /2} \xi^{|\ell|/2} W(\xi).
\end{equation} 

It is instructive to note that in case of the standard Landau problem for free electrons in a uniform magnetic field the solution (\ref{5}) is obtained taking $ r \rightarrow \infty$ (which is to say $\xi \rightarrow \infty$). On the other hand, in our system we consider the particles confined in an atomic cloud which implies that the value of $r$ is limited. Meanwhile, we can obtain an analytical solution in the form of Eq. (\ref{5}), in the limit \cite{20,23}
\begin{equation}\label{}
\frac{m \omega}{2 \hbar} \gg 1,
\end{equation} 
which leads to the following restriction on the effective magnetic field strength
\begin{equation}\label{cond1}
B_{\mathrm{eff}} \gg 2 \hbar.
\end{equation}

By substituting solution (\ref{5}) into Eq. (\ref{4}), the ones have
\begin{eqnarray}\label{}
\xi \frac{d^2}{d\xi^2} W(\xi) + \bigg[|\ell|  + 1 && -  \lambda \xi  \bigg] \frac{d}{d\xi} W(\xi)  + \bigg[ \beta - \lambda \frac{\left(|\ell| +1\right)}{2} \nonumber\\ &&+ \left( \frac{\lambda^{2}}{4} - \frac{\lambda}{4}\right)\xi \bigg] W(\xi) = 0.
\end{eqnarray} 

In the limit 
\begin{equation}\label{limit}
\lambda \rightarrow 1,
\end{equation}
i.e., $\alpha B_{0}^2 \gg m $, (Note that, in principle, this limit would be characterized by a high magnetic field), one arrives in the confluent hypergeometric equation 
\begin{eqnarray}\label{}
\xi \frac{d^2}{d\xi^2} W(\xi) + \bigg[|\ell|  + 1 && -  \xi  \bigg] \frac{d}{d\xi} W(\xi)\nonumber\\ &&+ \bigg[ \beta  - \frac{\left(|\ell| +1\right)}{2} \bigg] W(\xi) = 0,
\end{eqnarray} 
which is satisfied by the confluent hypergeometric function
\begin{equation}\label{confluent}
W(\xi) =  F \left(-\beta + \frac{\left(|\ell| +1\right)}{2}, |\ell| +1, \xi \right).
\end{equation} 

In order to have normalization of the wave function, the series in (\ref{confluent}) must be a polynomial of degree $\nu$, therefore,
\begin{equation}
\beta - \frac{\left(|\ell| +1\right)}{2} = n_{\xi},
\end{equation}
where $n_{\xi}$ is an integer number. From this condition, we obtain the energy eigenvalues 
\begin{equation}
E_{n_{\xi}, \ell} =  \hbar \omega \left(n_{\xi}  + \frac{|\ell|}{2} + \frac{\ell}{2}  + \frac{1}{2} \right).
\end{equation} 

It is useful to introduce a new quantum number 
\begin{equation}\label{}
n =  n_{\xi}  + \frac{|\ell| + \ell}{2}.
\end{equation} 

Then the energy spectrum acquire the standard form of the Landau spectrum
\begin{equation}\label{landau}
E_{n} =  \hbar \omega \left(n + \frac{1}{2} \right),
\end{equation} 
where $ n = 0,1,2,... $.

The energy levels in Eq.(\ref{landau}) are highly degenerate and are spaced by $\frac{\hbar\alpha \rho_{0} B_{0}}{m \epsilon_{0}} $. The effective magnetic length is $ l = \sqrt{\frac{\hbar \epsilon_{0}}{\alpha B_{0} \rho_{0}}} $ and the degeneracy is
\begin{equation}\label{degeneracy}
D =  \rho B_\mathrm{eff},
\end{equation} 
where $ \rho = \frac{A}{ h} $ and $ B_\mathrm{eff} = \frac{\alpha B_{0} \rho_{0}}{\epsilon_{0}}$. Note that the degeneracy linearly depends on the magnetic field $B_\mathrm{eff}$ and it is limited by the fact that we have consider a finite atomic trap.

\section{Magnetic oscillations for neutral particles subject to an electromagnetic field}

\begin{figure}[h!]
        \centering
       \includegraphics[scale=0.5]{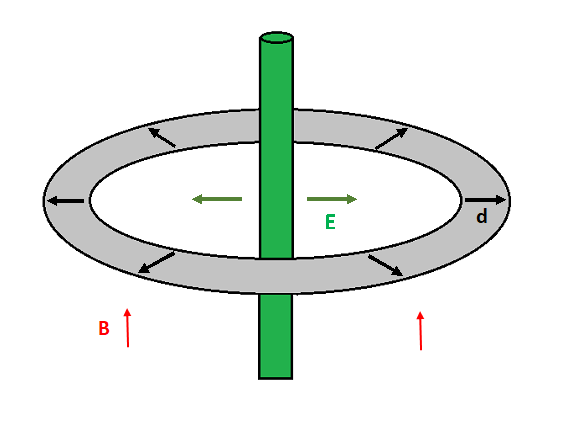}       
      \caption{Experimental arrangement proposed to observation of the atomic analog de Haas van Alphen effect. The electric dipole moment, {\bf d} (denoted by the black arrows), of the atoms in a 2D trap are aligned, via an infinitely long
charged wire (green tube), radially and reside in a uniform magnetic field perpendicular {\bf B} (red arrows) to the plane.}
\label{setup}
\end{figure}

For the realization of the atomic analog dHvA oscillations we propose a system composed by ultracold Rydberg atomic cloud confined in a toroidal trap with the dipoles aligned radially, via an electric field generated from a line of charge passing through the toroid, and a uniform external magnetic field is aligned perpendicularly to the toroid Fig.(\ref{setup}). 

The use of the ultracold Rydberg atoms is justified by the extreme properties of the highly excited atoms compared to atoms in ground state, such as very high dipole polarizabilities, magnetic moments and atom-atom strengths. Such properties turn the Rydberg atoms very sensitive to electric and magnetic fields  \cite{24,25,26}. This leads to a stronger HMW interaction for the Rydberg atom in presence of an electromagnetic field, as compared to an atom in the ground state.

In what follows, we consider a 2D cold Rydberg atomic cloud with an area of $ A \sim 150 \ \ \mathrm{\mu m^{2}} $ and contains $ N \sim 10^{4} $ atoms of $ ^{87}\mathrm{Rb}  $ \cite{27,28}. In addition, we assume $\alpha \sim 10^{9} \ \ \mathrm{Hz/(V/cm)^{2}} $ \cite{24}. Using these experimental parameters the condition (\ref{cond1}) becomes $|B_\mathrm{eff}| \gg 2.108 \times 10^{-34} \ \ \mathrm{T}$, and the degeneracy can be expressed as
\begin{equation}\label{}
D = 2.26 \times 10^{23}  B_\mathrm{eff}.
\end{equation}

Note that, the lowest landau level regime of the system $ D = 10^{4} $ is achieve when $ B_\mathrm{eff} = 4.42\times 10^{-20} \ \ \mathrm{T} $.

Henceforth, we provide a simple qualitative picture of dHvA oscillations. We consider the system at zero-temperature limit and containing a fixed number of $ N $ atoms. We do not take into account the temperature smearing of the quantum oscillations. For a given value of the artificial magnetic field $B_\mathrm{eff}$, we assume that the lowest $p$ landau levels (where $p$ is a positive integer) are completely filled with $ p D $ atoms each one and the highest $ (p+1)\mathrm{th} $ level is partly occupied with $ N - p D $ atoms. In this case, the Fermi level lies in the $ (p+1)\mathrm{th} $ level. From Eq. (\ref{degeneracy}) we can see that the degeneracy degree $D$ of the system decreases when the effective magnetic field is decreased. As a consequence, fewer atoms can be accommodated on each Landau level and the atomic population of the highest $ (p+1)\mathrm{th} $ energy level will range from completely full to entirely empty. This decrease by 1 in the number of Landau levels occupied is the origin of the dHvA oscillations.

\begin{figure}[!ht]
\centering
\includegraphics[scale= 0.85]{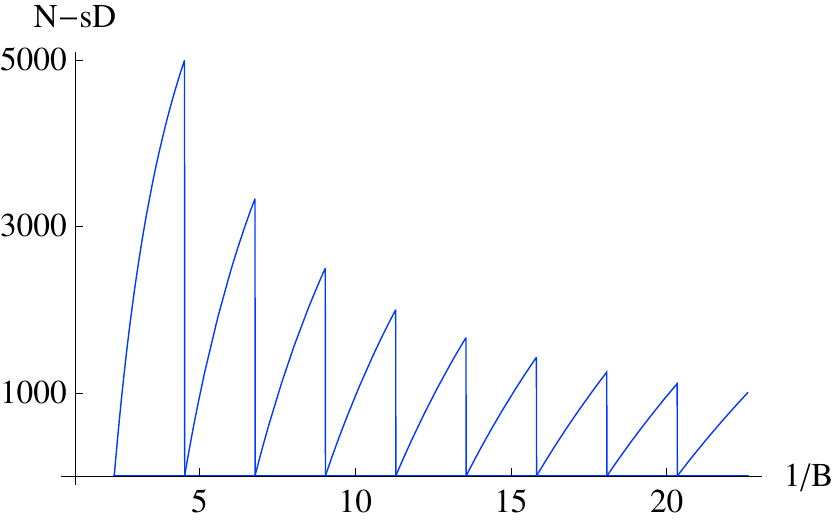}
\caption{(Color online) Number of $ N - p D $ atoms in the Landau level partially filled as a function of the reciprocal magnetic field (in units of $ 10^{20}  \ \mathrm{T_{eff}^{-1}} $).} 
\label{Number_atom}
\end{figure}

Fig.(\ref{Number_atom}) displays how the population of the highest partly occupied level of a 2D ultracold cloud with $ N = 10^{4} $ $ ^{87}\mathrm{Rb}  $ atoms when the magnetic field is slept in the range $ 4.42\times 10^{-20} \ \ \mathrm{T_{eff}} \leq {B_\mathrm{eff}} \leq   4.42\times 10^{-21} \ \ \mathrm{T_{eff}} $ or equivalently $ 2.26 \times 10^{19} \ \mathrm{T_{eff}^{-1}} \leq \frac{1}{B_\mathrm{eff}} \leq   2.26 \times 10^{20} \ \mathrm{T_{eff}^{-1}} $. At $ \frac{1}{B_\mathrm{eff}} = 2.26 \times 10^{19} \ \mathrm{T_{eff}^{-1}} $ only the lowest level $ p = 0 $ is populated with $ 10^{4} $  atoms and the upper level $ (p+1)\mathrm{th} = 1 $ is empty. As the reciprocal magnetic field is increased the level $ p = 1 $ starts to accommodate atoms until it is fully occupied with $ 5 \times 10^{3} $ atoms. Then a new upper level $ (p+1) \mathrm{th} = 2 $ becomes populated and so on. Note that singularities appear at strength of the magnetic field where a new energy level becomes occupied. The distance between such singularities are regularly spaced and defines the dHvA period $ \Delta(\frac{1}{B_\mathrm{eff}}) = \frac{\rho}{N} = 2.26 \times 10^{19}  \ \mathrm{T_{eff}^{-1}} $. It is this jump of atoms to a higher energy level that causes the oscillations in the energy and in the effective magnetization of the atomic gas as a function of the inverse magnetic field. 

The energy of the neutral atoms in the partly filled level is given by
\begin{equation}\label{energy3}
 \varepsilon = (N - p D)\hbar\omega(p+\frac{1}{2})
\end{equation}
which can be rewritten in a most appropriate form as
\begin{equation}\label{energy4}
\varepsilon = - \mu^{\mathrm{eff}}_{B}\rho\left[ p B_\mathrm{eff}  - \frac{N}{\rho}\right] \left[(p+1) B_\mathrm{eff} - \frac{N}{\rho} \right],
\end{equation}
for $\frac{\rho s}{N} < \frac{1}{B_\mathrm{eff}} < \frac{\rho (s + 1)}{N} $. Here $ \mu^{\mathrm{eff}}_{B} = \frac{\hbar}{m} = \frac{\hbar}{M+\alpha B_{0}^{2}} $ is an effective Bohr magneton. 

\begin{figure}[h!]
        \centering
        \includegraphics[scale=0.85]{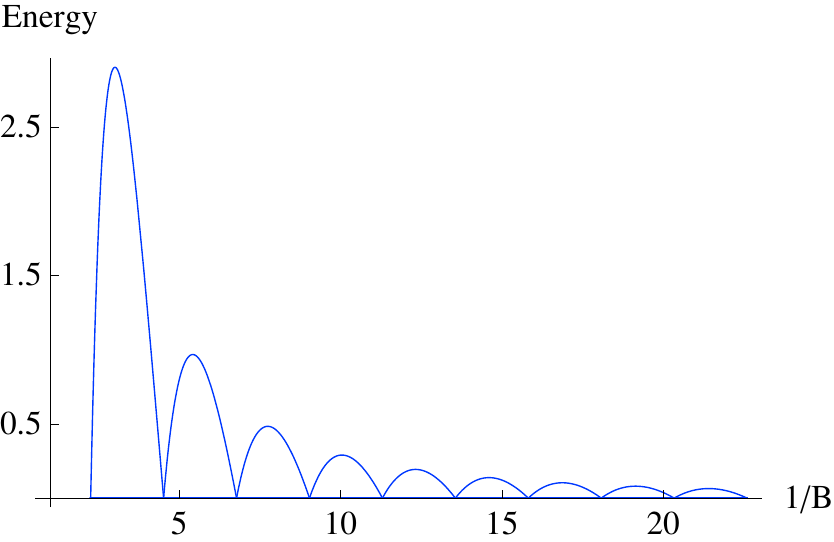}       
        \caption{
		Variation of the energy (in units of $ 10^{-56}\ \ \mathrm{J} $) of the partially occupied Landau level with respect to the inverse effective magnetic field (in units of $ 10^{20}\ \ \mathrm{T_{eff}^{-1}} $). For calculations, we take $ M_{\mathrm{Rb}} = 1.443\times10^{-25}\ \mathrm{kg} $.
}
\label{energy}
\end{figure}

As Fig. (\ref{energy}) shows, the energy $\varepsilon$ oscillates with a quadratic dependence on each dHvA period.  Note that if we set either $ \frac{1}{B_\mathrm{eff}^{p}} = \frac{p\rho}{N} $ or $ \frac{1}{B_\mathrm{eff}^{p+1}} = \frac{(p+1)\rho}{N} $ there is no Landau level partly populated and the Eq. (\ref{energy4}) is zero. On the other hand, at $ \frac{1}{B_\mathrm{eff}} = \frac{\rho}{N}\frac{p(p+1)}{p+\frac{1}{2}} $ the energy $\varepsilon$ has a local maximum. The amplitude of the oscillations falls because a new higher level passes to be occupied each time with fewer particles as $ \frac{1}{B_\mathrm{eff}} $ varies.

\begin{figure}[h!]
        \centering
        \includegraphics[scale=0.85]{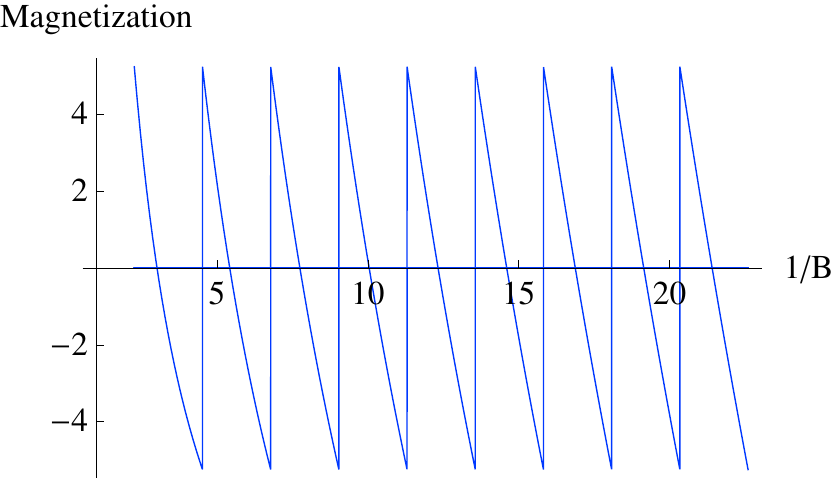}       
        \caption{
		Effective Magnetization (in units of $ 10^{-36}\ \mathrm{\frac{J \cdot  s}{T}} $) as a function of the inverse magnetic field (in units of $ 10^{20}\ \ \mathrm{T^{-1}} $).
}
\label{magnetization}
\end{figure}

The effective magnetization $\mathcal{M} $ is obtained taking $ - \frac{\partial \varepsilon}{\partial B_\mathrm{eff} } $, then we have 
\begin{eqnarray}\label{Magnetization}
\mathcal{M} =  \mu^{\mathrm{eff}}_{B}\rho\left[ 2B_\mathrm{eff} p(p+1) - \frac{N}{\rho} (2p+1) \right], 
\end{eqnarray}
where $ pD < N \leq (p+1)D $.
In Fig. (\ref{magnetization}) we display the dHvA oscillations of the synthetic magnetization as a function of the inverse magnetic field calculated from Eq. (\ref{magnetization}). As expected, at $T = 0$ these oscillations have a saw-tooth shape with a constant amplitude. $\mathcal{M}$ is linear in $ \frac{1}{B_\mathrm{eff}} $ except in points for which a new energy level starts to be filled. In these points the Fermi level makes an abrupt jump between two Landau levels and consequently $\mathcal{M}$ experiences jumps of $ 2 N \mu^{\mathrm{eff}}_{B} \approx 2.11 \times 10^{-39} \ \mathrm{\frac{J \cdot s}{T}} $ at the end of each dHvA period. The effective magnetization is $ - N \mu^{\mathrm{eff}}_{B} $ (Landau diamagnetism) when just the first $ p $ energy levels are occupied, and it jumps to $ + N \mu^{\mathrm{eff}}_{B} $ as the $ (p+1)$th level starts to be populated, returning smoothly to $ - N \mu^{\mathrm{eff}}_{B} $ again when this new level is completely populated. 

It is important to observe that since the magnetic field in Eq. (\ref{Magnetization}) is artificial, the effective magnetization $\mathcal{M}$ is not directly observable. However, as discussed in \cite{29} there are indications that the dHvA oscillations should be observed in the angular momentum $\langle L_{z} \rangle $ of the atomic gas, which is directly analogous to the magnetization in the solid-state context.

From the period of the dHvA oscillations, we can obtain the area $S$ of the extremal cross-section of the Fermi surface of the atomic cloud through the Onsager-like relation
\begin{eqnarray}
S = \frac{2\pi N}{\hbar \rho} \nonumber
\end{eqnarray}

Using the physical parameters of the system we estimate the area of the Fermi circle in $ S \approx 2.64 \times 10^{15} \ \mathrm{m^{-2}} $.

\section{Conclusion}

In conclusion, we have proposed an experimental scheme to study dHvA oscillations for a cloud of a 2D neutral atomic gas. Under a certain field-dipole configuration the HMW interaction creates an effective magnetic field which leads the atoms organize into Landau levels. We have showed that the finite dimensions of the atomic cloud limit the artificial magnetic field strength. We have displayed quantum oscillations in the energy and effective magnetization of the gas as a function of the reciprocal magnetic field. We have estimated the area of the Fermi circle of the atomic cloud. The present scheme provides an applicable way to realization of the dHvA effect in ultracold neutral atoms.

\section{Acknowledgments}

We would like to thank CNPq, CAPES and FAPESQ for financial support.

\end{document}